\documentclass[twoside,slac_one]{revtex4}
\usepackage{graphicx}
\usepackage{fancyhdr}
\usepackage{amsmath} 
\usepackage{bm}
\usepackage{amsxtra}
\usepackage{amssymb}
\usepackage{amsthm}
\usepackage{latexsym}
\usepackage{lscape}

\begin{document}

\title{Hollow cone sieve for top}

%

\author{Vernon Barger, Peisi Huang}
\affiliation{Department of Physics, University of Wisconsin, Madison, WI 53706, USA}

\begin{abstract}
A method of top tagging is introduced.  Using the anti-kt algorithm to define jets, events with $ n_j$  = 2 fat jets of cone size R = 1.5  are decomposed into R = 0.6 sub-jets and retained if $n_j$ (R= 0.6) $\geq$ 4 .  One pair of sub-jets reconstructs the W-mass and another jet is tagged as a b-jet,  as necessary for hadronic or semileptonic events of $t\overline{t}$ origin.   This 'hollow cone'  method distinguishes the $t\overline{t}$ events from the light parton QCD dijet events. Our simulations are made for the LHC at 7 TeV.
\end{abstract}

\maketitle

\thispagestyle{fancy}


\section{Introduction}
The top quark is important in the Standard Model (SM)  because of its the large coupling to the Higgs field. Also, top is of great interest in new physics extensions of the SM because one or more new particles commonly decay into a single top quark or into top-quark pairs.  Examples include SUSY \cite{Dimopoulos:1981zb}, where the stop companion of top may decay to top  and the gluino companion of the gluon may decay to top pairs, the Little Higgs Theory \cite{ArkaniHamed:2001nc} where a heavy top may decay to top, and the Randall-Sundrum model \cite{Randall:1999ee} where the decays of Kaluza-Klein gluons may lead to an enhancement of top signals.   Thus, the study of top at the Tevatron and at the Large Hadron Collider (LHC) should be an excellent probe of new physics.   We focus on the semileptonic channel (one t$\rightarrow$ b l $\nu$ decay and one t$\rightarrow$b q $\mathrm{\overline{q}}$ decay) and the hadronic channel (both tops decay to quarks); the leptonic channel with both tops decaying via t $\rightarrow$b l $\nu$ has a clean signal but a reduced rate.  The expected $t\overline{t}$ event ratios in the SM are hadronic : semileptonic : leptonic = 3.2 : 1: 0.32 .  Silicon vertex detectors have enabled the tagging of b-events with high efficiency ($\sim$ 50$\%$) using the secodary vertes algorithm \cite{Abazov:2008gc} at the Tevatron and b-tagging efficiencies of  60$\%$ \cite{Aad:2009wy}, 65$\%$ \cite{Volpe:2009ns} are estimated for the ATLAS and CMS detectors.

At the Tevatron collider, b-tagging and kinematical selection have been used to identify $t\overline{t}$ events in the fully hadronic channel \cite{Wicke:2004cg}.  For the semi-leptonic channel, the CDF collaboration employs b-tagging, jet multiplicity and an isolated electron or muon to do the event selection, and uses a matrix element method to tag a top \cite{Aaltonen:2010yz}. The D0 collaboration selects events with exactly 4 jets and assigns the jets : 1 jet associated with the lepton and the other 3 jets with the hadronic top decay in a way that minimizes the reconstructed mass difference between the two top quarks \cite{Abachi:1995iq}.\\

The LHC is a top factory: in the SM about 8,000 top pairs should have been produced with more than 47 {$\mathrm{pb^{-1}}$ integrated luminosity already taken per detector at 7 TeV. Since the LHC center-of-mass energy is high compared to the top mass, the tops will typically be highly boosted, so that the decay products are close to each other, as illustrated in Figure \ref{fig:FatJet}.
\begin{figure}[here]
\includegraphics[width = 0.8\textwidth]{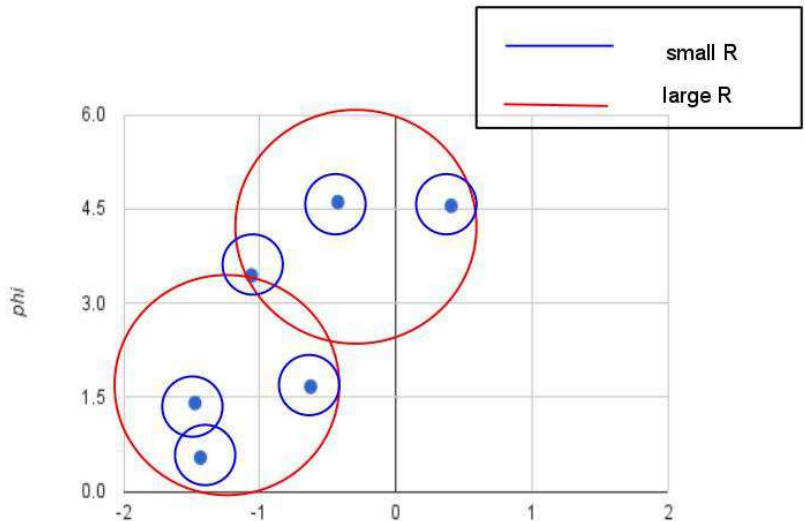}
\caption{Subjets of a top in $\eta$ $\phi$ plane}
\label{fig:FatJet}
\end{figure}
Thus, in the detector, at first sight, the top decay products may look like a fat jet instead of the several separate ones of which it is composed. Butterworth $et$  $al.$ look into subjets inside a fat jet \cite{Butterworth:2008iy} \cite{Butterworth:2002tt}, to tag a Higgs or a W boson. One issue arises with the fat jet is that soft QCD contamination becomes relevant in the analysis of the jet substructure. Several techniques have been developed to deal with that complication. Jet filtering \cite{Butterworth:2008iy} keeps only several hard subjets, while jet trimming \cite{Krohn:2009th} strives to eliminate the soft ones. Jet pruning \cite{Ellis:2009su} avoids soft and large angle jet recombinations. Following these techniques, there are several algorithms to tag a top. Kaplan $et$ $ al.$ \cite{Kaplan:2008ie} use the ratio of transverse momentum of the subjet to that of the original jet (the cut is 0.1 or 0.05) to find out which subjet is a $``$ hard subjet $"$ and which is from soft radiation, then apply a cut on invariant masses of the fat jet and 2 of the subjets, requiring them to be within the mass windows of top and W respectively. Also, a cut on helicity angle of the W, which is the angle between the reconstructed top and one of the W decay products in the reconstructed W rest frame \cite{Chwalek:2007pc}, can be applied. Plehn $et$ $al.$ \cite{Plehn:2009rk} use a mass drop criteria to find sub-jets (discussed below ) and use a fixed mass window for top and W reconstruction, to tag tops with $p_{T}$ above 200 GeV. There are also several top tagging algorithms using other kinematics variables, without tagging a W or a b, like two-body or multi-body kinematics \cite{Thaler:2008ju}, separation between lepton and hadronic activities of top \cite{Thaler:2008ju} \cite{Rehermann:2010vq}, and jet mass distribution \cite{Almeida:2008tp}.\\
\section{Search strategy}
In the top reconstruction, to catch all the three main decay products of a top as a single fat jet, it is natural to use a large jet size, in other words, a large R parameter in the jet clustering, otherwise, several seperate jets will be constructed instead of one. Here R is the jet size defined by R = $\mathrm{\sqrt{\eta^2+\phi^2}}$ where $\eta$ is the pseudo-rapidity difference of two particles and $\phi$ is their azimuthal angle difference.  The lower value of R = 1.5 for the top jet is suggested by the the analyses of ref. \cite{Plehn:2010st}. Now consider a top pair. If a large R = 1.5 is used, it is likely that two jets will be constructed in the event (one from the top and the other from the $\mathrm{\overline{t}}$), while more will be constructed using a small R. The light jets behave differently from the top jets, in that the number of reconstructed light jets does not vary with R. So, after subtracting light jets from dijet events, the top contribution can be seen in the variation of the number of jets with cone size R.
\begin{figure}[here]
\includegraphics[width = 0.45\textwidth]{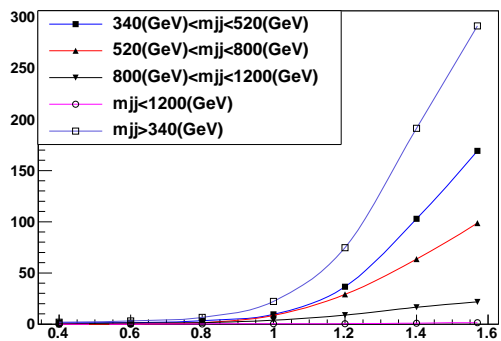}
\caption{Number of top pairs in dijet events}
\label{fig:TopVsR}
\end{figure}
Figure \ref{fig:TopVsR} shows number of jets versus cone size R in events first selected as dijets with  R = 0.6.  The number of top pair events increases with R.   Results are for the LHC at 7 TeV with 50 $\mathrm{pb^{-1}}$.
We develop the $``$hollow cone $"$ idea to tag top pairs. Consider the anti-kt algorithm as a $``$ perfect cone $"$ algorithm. When a larger cone size is used, both a $t\overline{t}$ event and a QCD dijet event will give two jets, when a smaller cone size is used, a $t\overline{t}$ event will have more jets while a QCD dijet event still have two. This means, for a fat jet with a large cone size, after subtracting a jet of small cone size in the interior, if some jets remain in the hollow cone, the jet is likely to be a top jet, and if there is no jet in the hollow cone, it is likely to be a light quark jet or a gluon jet.

 Our top tagging algorithm proceeds in the following steps, trying to separate top pairs from the QCD dijet events:\\
1$)$ Reconstruct jets using the anti-kt \cite{Cacciari:2008gp} jet algorithm with R = 1.5 to obtain a set of jets. The number of jets is $n_{jets}$. \\
2$)$ Redo the jet reconstruction,with R = 0.6 (or R = 0.7), following recent works of ATLAS \cite{Collaboration:2010eza} and CMS \cite{Khachatryan:2010jd}, to obtain another set of jets.\\
3$)$ Keep the event as a $t\overline{t}$ candidate if $n_{jets, R=1.5 }$ $=$ 2 and $n_{R = 0.6}$ $>$2 . \\
4$)$ Go into the 2 jets reconstructed in step 1, find all the subjets for each fat jet, using the method described in \cite{Plehn:2009rk}, as follows. For a fat jet of invariant mass of $m_j$, undo the last step of jet clustering to obtain two jets $j_1$ and $j_2$, with invariant masses $m_{j1}$ and $m_{j2}$ ( $m_{j1} > $  $m_{j2}$ ). If $m_{j1}$ $< $0.9 $m_j$, keep both $j_1$ and $j_2$, otherwise, keep only $j_1$ to add to the subjet list and decompose further. Add $j_i$ to the jet substructure list if $m_{ji}$ $<$ 30 GeV, otherwise decompose $j_i$ iteratively. If the total number of subjets is less than 4, reject the event, because a hadronic top and one semileptonic top should give 4 subjets in total, and two hadronic tops will result in 6 subjets. \\
5$)$ See whether there is a W inside either of the 2 fat jets, if not, reject the event. To do this, look into a fat jet and iterate over all of the 2 subjets configurations. After the jet filtering \cite{Butterworth:2008tr}, if the invariant mass of the 2 subjets falls in the window of 65 GeV to 95 GeV, tag that configuration as a W. A similar method of W-tagging is discussed in \cite{Kaplan:2008ie}, \cite{Plehn:2009rk}, \cite{Cui:2010km},  and \cite {CMS-PAS-JME-09-001}.\\
6$)$ See whether either of the 2 jets has a subjet can be tagged as a b jet.The jet candidates of a W must not be tagged as a b-jet. Keep other b-tagged events.\\
7$)$Any event that survives the above sequence is tagged as a $t\overline{t}$ event.\\
Our method is only applicable to $t\overline{t}$ pair production without extra jets, but this requirement only reduces the cross section from 160 pb to 100 pb.\\

\section{Signal and Background} The main backgrounds here are $Wb\overline{b}$ and $Zb\overline{b}$. Since there will be b jets in both cases, and the Z mass is close to W mass, these two backgrounds are indistinguishable in their hadronic decay channels. Other backgrounds are the QCD dijets from light quarks and gluons and QCD multi-jet events. QCD dijets events will be gotten rid of by the $``$ hollow cone $"$ cut. Since the anti-kt jet algorithm is collinear safe and infrared safe, the QCD dijet background can be reliably removed.  The QCD trijet background can be eliminated by the number of subjets. With fake b-jets, Wjj and Zjj events contribute to the background also. 

We generate the parton level events with MadEvent\cite{Alwall:2007st}, then use Pythia\cite{Sjostrand:2006za} to do the parton shower and hadronization. FastJet\cite{Cacciari:2005hq} is used to reconstruct jets and analyze the jet substructure. Then we apply the following cuts in sequence : \\
cut 1 : The $``$hollow cone$"$ sieve. Require $n_{jets}$ $=$ 2 and $n_{veto}$ $>2$.\\
cut 2 : Total number of subjets $\geq$ 4.\\
cut 3 : A hadronic W can be tagged.\\
cut 4 : A b jet can be tagged. \\
The cut flow table of signal and background is in Table \ref{table:CutFlow}. We assume a 0.5  b-tagging efficiency and a light jet rejection of 1/200\cite{Aad:2009wy} here.
\begin{table}[h]
\centering
 \begin{tabular}{c|c|c|c|c|c}
\hline \hline
& cross section($\mathrm{pb}$)& cut 1($\mathrm{pb}$)& cut 2($\mathrm{pb}$)& cut 3($\mathrm{pb}$)& cut 4($\mathrm{pb}$) \\
\hline
$t\overline{t}$& 100.00& 12.63& 7.59&5.39&4.05\\
$Wb\overline{b}$& 239.52 & 63.93 & 1.40 & 0.20&0.18\\
$Zb\overline{b}$ &124.81&23.55&1.20&0.57&0.43 \\
$Wjj$& 2458 &771.4 &91.9&8.00&0.08 \\
$Zjj$&7727.5 & 478.3 & 121.3 & 25.5 & 0.26 \\ 
\hline
\end{tabular}
\caption{Cut flow table for signal and backgrounds.}
\label{table:CutFlow}
\end{table}
The reconstructed W mass and top mass can be seen in figure \ref{fig:Wmass} and \ref{fig:top_mass},respectively.\\
\begin{figure}[here]
\includegraphics[width = 0.45 \textwidth]{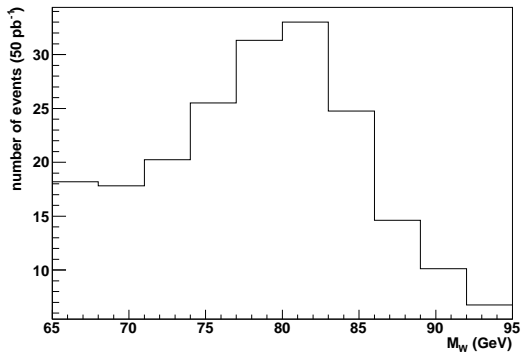}
\caption{Reconstructed W mass}
\label{fig:Wmass}
\end{figure}
\begin{figure}[here]
\includegraphics[width = 0.45\textwidth] {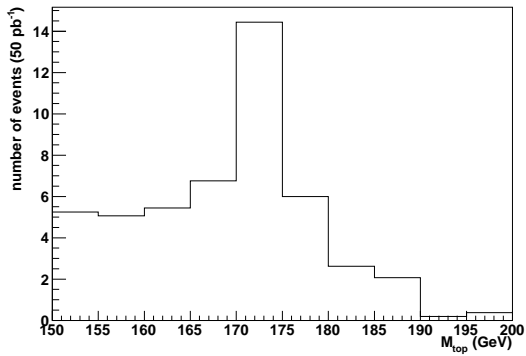}
\caption{Reconstructed top mass}
\label{fig:top_mass}
\end{figure}
The resulting ratio of hadronic tops to semileptonic tops is 2.81, which is consistent with the ratio of decay branching fractions of 3.13.
The transverse momentum distribution of the tagged t and $\mathrm{\overline{t}}$ is shown in figure \ref{fig:pt} and compared with the transverse momentum of the QCD dijet. It shows that the method is picking top jets instead of light jets, and also demonstrates that top jets with relatively low $\mathrm{p_t}$ can be tagged.  Former top tagging techniques require the p$_T$ of the top to be harder than 200 GeV \cite{Plehn:2009rk}.\\
\begin{figure}[here]
\includegraphics[width = 0.45 \textwidth]{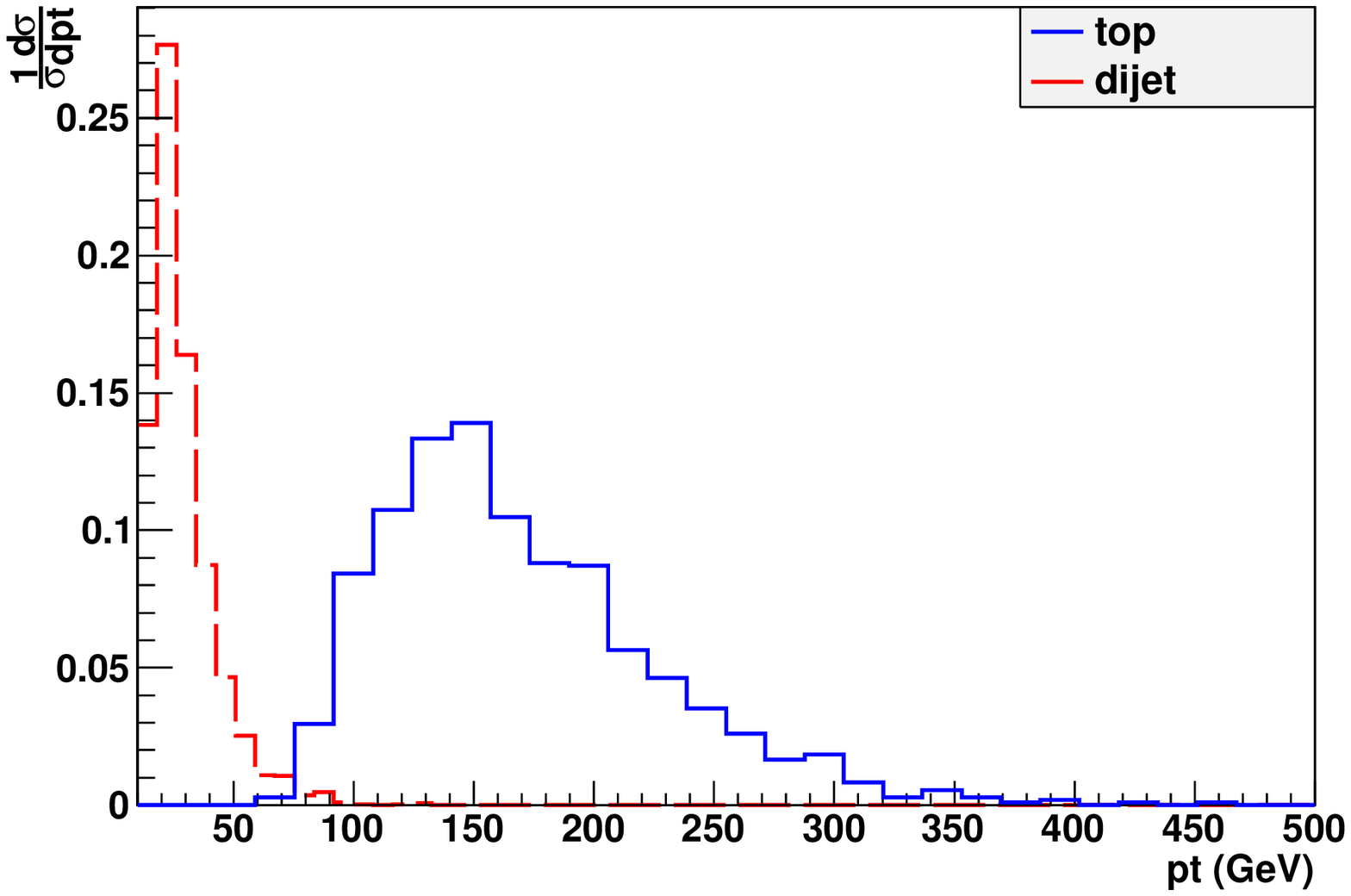}
\caption{Normalized $p_T$ distributions for top and dijets}
\label{fig:pt}
\end{figure}
Figures \ref{fig:dr_t} and \ref{fig:dr_w} show the distance in R of W with the nearest b jet for $t\overline{t}$ and $Wb\overline{b}$. The $t\overline{t}$ events are more likely to have a b jet close to the W than in $Wb\overline{b}$ jets. In the $t\overline{t}$ events, the first peak comes from the b jets associated with the W in a top candidate. \\
\begin{figure}[here]
\includegraphics[width = 0.45 \textwidth] {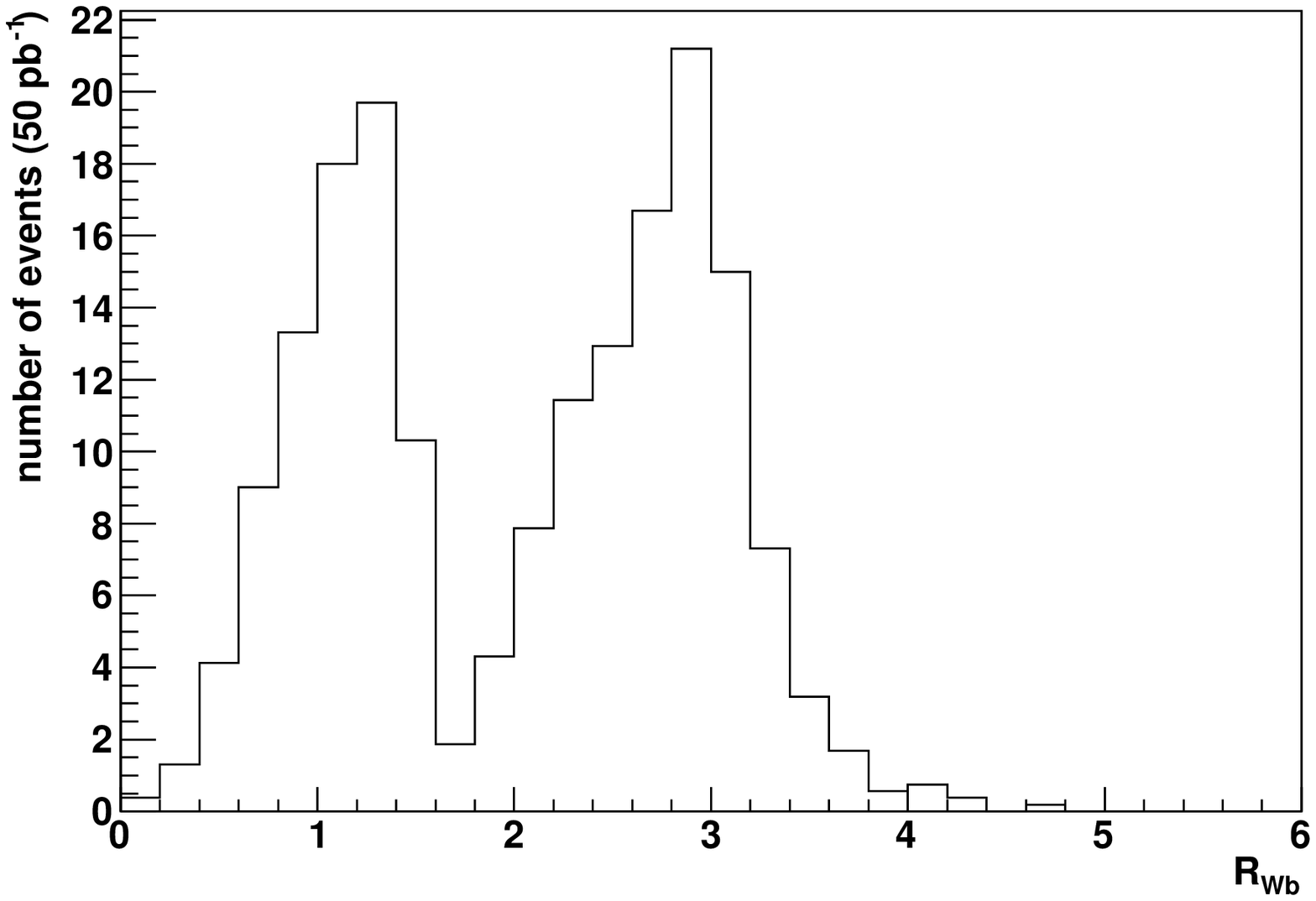}
\caption{Distance in R between the W-boson and the nearest b jet  in $t\overline{t}$} 
\label{fig:dr_t}
\end{figure}
\begin{figure}[here]
\includegraphics[width = 0.45\textwidth]{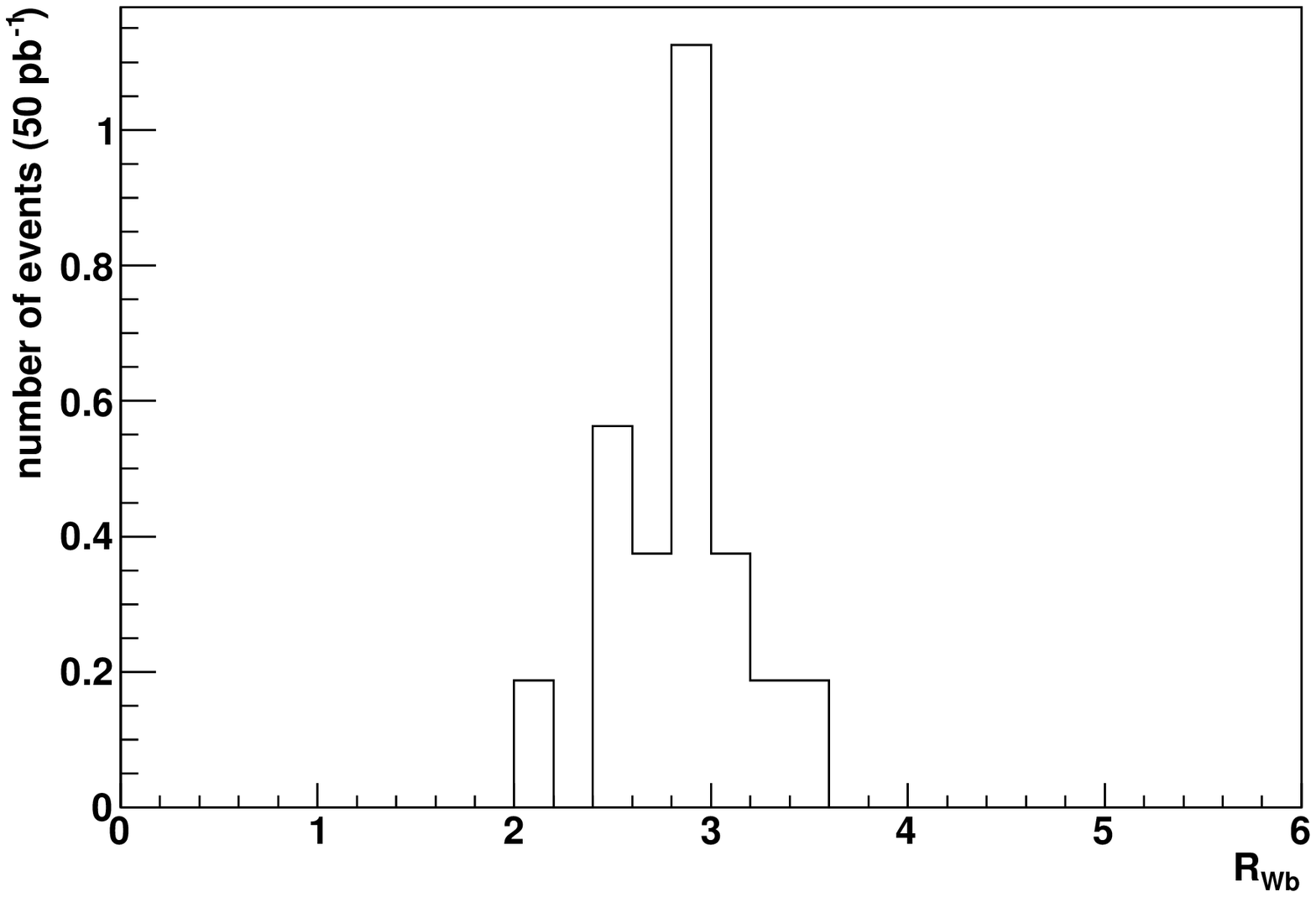}
\caption{Distance in R between the W and the nearest b jet in $Wb\overline{b}$} events.
\label{fig:dr_w}
\end{figure}
\section{Outlook}
 This method tags 4050 $t\overline{t}$ events at 7 TeV,while the method described in \cite{Plehn:2009rk} tags 1457 events at 14 TeV . The sieve method can be also used in identifying new physics that has a top in the final state. The tagged top can serve as a distinguishable signature. Also, the $``$ hollow cone $"$ can be used for discovering new, relatively heavy and boosted particles at the LHC after taking the SM top contribution into consideration. If the new particle has a mass that is not very close to the top mass, the mass of the new particle can be reconstructed by the invariant mass of the filtered subjets.

\begin{acknowledgments}
This work was supported in part by the U.S. Department of Energy under grant No. DE-FG02-95ER40896.
\end{acknowledgments}

\bigskip 
\bibliographystyle{unsrt}
\bibliography{toprefs}

\end{document}